\documentclass[a4paper,conference]{IEEEtran}
\IEEEoverridecommandlockouts
\usepackage{graphicx,graphics,epsfig,epstopdf,float}
\usepackage{amsmath} 
\usepackage{amssymb}  
\usepackage{mathrsfs}
\usepackage{enumerate}
\usepackage{subfigure}
\usepackage{bm}
\usepackage{psfrag}
\usepackage{cite}
\usepackage{mdwlist}
\usepackage{longtable}
\usepackage{array}
\usepackage{arydshln}
\usepackage{acronym}  
\usepackage{verbatim}

\usepackage{diagbox}
\usepackage{bm}
\usepackage{multirow}
\usepackage{flushend}

\newcommand{\V}[1]{\bm{#1}} 
\newcommand{\M}[1]{\bm{#1}} 

\usepackage{booktabs}
\usepackage{cite}
\usepackage{amsmath,amssymb,amsfonts}
\usepackage{algorithm}
\usepackage{algpseudocode}
\usepackage{graphicx}
\usepackage{textcomp}
\usepackage{xcolor}
\def\BibTeX{{\rm B\kern-.05em{\sc i\kern-.025em b}\kern-.08em
    T\kern-.1667em\lower.7ex\hbox{E}\kern-.125emX}}
\begin{document}

\title{Cramer-Rao Lower Bound Analysis for OTFS and OFDM Modulation Systems\\
}

\author{\IEEEauthorblockN{Bowen Wang, Jianchi Zhu, Xiaoming She, Peng Chen}
\IEEEauthorblockA{\textit{China Telecom Research Institute} \\
\textit{Beijing, China}\\
Email: \{wangbw2, zhujc, shexm, chenpeng11\}@chinatelecom.cn}
}

\maketitle

\begin{abstract}
The orthogonal time frequency space (OTFS) modulation as a promising signal representation attracts growing interest for integrated sensing and communication (ISAC), yet its merits over orthogonal frequency division multiplexing (OFDM) remain controversial. This paper devotes to a comprehensive comparison of OTFS and OFDM for sensing from the perspective of Cram$\acute{\text{e}}$r-Rao lower bounds (CRLB) analysis. To this end, we develop the cyclic prefix (CP)-Free and CP-added model for OFDM, while for OTFS, we consider the Zak transform based and the Two-Step conversion based models, respectively. Then we rephrase these four models into a unified matrix format to derive the CRLB of the delays and doppler shifts for multipath scenario. Numerical results demonstrate the superiority of OTFS modulation for sensing, and the effect of physical parameters for performance achievement.
\end{abstract}


\section{Introduction}
\indent The integrated sensing and communication (ISAC) has emerged as a key technology in 6G due to the capability of functioning both as radars and communication systems\cite{LiuHuaLi:J22}. One of the most important topics of ISAC is the integrated waveform design.

The well-known Orthogonal Frequency Division Multiplexing (OFDM) has achieved great success in wireless communications and has been extensively studied for localization\cite{StuBarMcl:J04,WanSheMaz:J14}. However, its spectral efficiency suffers from inter-symbol interference (ISI) and inter-carrier interference (ICI). Besides, the ability of its channel characterization is limited by the uncertainty principle, which may cause a deficiency of sensing ability\cite{StuWie:J11}. Recently, a new modulation scheme named orthogonal time frequency space (OTFS) was proposed, which multiplexes quadrature amplitude modulation symbols in the delay-Doppler (DD) representation, therefore holds promise for more accurate capture of the physics of the channel\cite{RavPhaHon:J18}. Motivated by its potentiality for ISAC, there emerge several research direct at OTFS based channel parameters estimation\cite{GauKobCai:J20,RavPhaHon:C19}. However, it still remains for us to exhibit the fundamental limit of OTFS based parameters estimation, especially the comparative analysis with OFDM. The Cram$\acute{\text{e}}$r-Rao lower bound (CRLB) as a fundamental limit on the variance of any unbiased estimator in statistical inference theory, is commonly used as a benchmark to evaluate various estimation algorithms and can be viewed as an effective criterion to guide practical algorithms design\cite{Sen:J93}.

Our analysis encompasses four modem schemes. The original OFDM transmission does not contain cyclic prefixes (CP), which we refer to as CP-Free OFDM. In order to resist the inter-symbol interference and inter-carrier interference caused by multipath propagation and frequency offset, respectively, a more standard CP-OFDM modulation which add a CP in front of each symbol is used in practical applications. As for OTFS, in order to ensure compatibility with existing OFDM systems, most prior work consider a two-step conversion on the receiver side, where the received time-domain signal is firstly converted to the time-frequency domain using an OFDM demodulator, followed by post-processing of this time-frequency signal into DD domain\cite{RavPhaHon:J18}. Recently, studies show that the direct conversion of the received time-domain signal to the DD domain via the Zak transform, is also feasible\cite{Moh:J21}. The resulting expression for the sampled DD domain signal model is not the same as that obtained by the two-step conversion. We refer to an OTFS modulation based on the direct conversion as the ¡°ZAK OTFS¡± and that based on the two-step conversion as the ¡°two-step OTFS¡±. We devote to clarifying the sensing abilities of OTFS and OFDM by deriving the CRLB of parameter estimation under these four signal models.

The remainder of the paper is organized as follows. In Section $\mathrm{II}$, we construct point-to-point transmission models for the four types of modem schemes and subsequently represented them as a formally unified matrix signal model. In section $\mathrm{III}$, we derive the CRLB of delay and Doppler shift estimations for the four types of schemes. Numerical results are provided in section $\mathrm{IV}$ to demonstrate the superiority of Zak-OTFS and Two-Steps OTFS, as well as the degradation of CP-OFDM for sensing.

\section{Signal Model}\label{sec2}
We respectively consider the OFDM and OTFS systems with single antenna transceiver over time-varying multipath channels. The physical resource includes bandwidth $B=M\Delta f$ and time duration $NT$, where $\Delta f$ and $T=1/\Delta f$ denote the subcarrier spacing (SCS) and the symbol duration, respectively. In this section, we establish the input-output models for the two systems considering four cases including CP-Free OFDM, CP-OFDM, Zak-OTFS, and Two-Steps OTFS.
\subsection{Point-to-Point CP-Free OFDM Transmission}
OFDM as an efficient multi-carrier modulation converts high speed serial communication data into $N$ parallel data streams, and modulates these data on $M$ orthogonal subcarriers, the composite OFDM baseband signal is then the summation of all the subcarriers
\begin{equation}
\begin{aligned}
f(t)=\sum_{m=0}^{M-1}\sum_{n=0}^{N-1}x[n,m]e^{j2\pi(f_c+m\Delta f)(t-nT)}g_{tx}(t-nT)
\end{aligned}
\end{equation}
where $f_c$ is the carrier frequency, $x[n,m]$ is the transmitted data on the $m$-th subcarrier of the $n$-th symbol duration. $g_{tx}(t)$ denotes the transmit pulse, and is set to be the rectangular window of duration $T$ by default.

Consider a time-frequency varying channel, i.e., its behavior changes with respect to (w.r.t.) the time instant and the subcarrier frequency, which may be represented as a continuum of elements that simultaneously provide both a corresponding delay and Doppler shift
\begin{equation}
\begin{aligned}
h(\tau,\nu)=\sum\nolimits_{p=1}^{P}h_p\delta(\tau-\tau_p)\delta(\nu-\nu_p)
\end{aligned}
\end{equation}
where $h(\tau,\nu)$ is the DD spread function, or interpreted as a continuum of nonmoving scintillating scatterers by the delay-spread functions which is the inverse Fourier transform of its spectrum and can be expressed as
\begin{equation}
\begin{aligned}
\gamma(t,\tau)=\int h(\tau,\nu)e^{j2\pi\nu t}d\nu
\end{aligned}
\end{equation}
then by ignoring the noise for the sake of simplicity, we can obtain the following input-output relationship as \cite{Bel:J63}
\begin{small}
\begin{equation}
\begin{aligned}
y'(t)=\int\gamma(t,\tau)f(t-\tau)d\tau=\sum\nolimits_{p=1}^{P}h_pf(t-\tau_p)e^{j2\pi\nu_pt}
\end{aligned}
\end{equation}
\end{small}%
where $h_p$, $\tau_p$, $\nu_p$ denote the channel gain, delay, and Doppler shift of the $p$-th path, respectively. By sampling $y'(t)$ every $T/M$ seconds in each OFDM symbol, we obtain
\begin{small}
\begin{equation}
\begin{aligned}
y&_{n',m'}^{'}=y'(t)|_{t=n'T+m'T/M}\\
=&\sum_{p=1}^{P}h_p\sum_{m=0}^{M-1}\!\sum_{n=0}^{N-1}\!x[n,m]e^{j2\pi m\Delta f((n'+\frac{m'}{M}-n)T-\tau_p)}\\
&g_{tx}(n'T+\frac{m'T}{M}-nT-\tau_p)e^{j2\pi\nu_p(n'T+\frac{m'T}{M})}\\
=&\left\{
    \begin{array}{lc}
        \!\!\!\sum_{p=1}^{P}h_p\sum_{m=0}^{M-1}x[n',m]\\
        e^{j2\pi m\Delta f(\frac{m'T}{M}-\tau_p)}e^{j2\pi\nu_p(n'T+\frac{m'T}{M})} & \!\!\mathrm{if}~ m'>\frac{M\tau_p}{T}, \\
        \!\!\!\sum_{p=1}^{P}h_p\sum_{m=0}^{M-1}x[n'-1,m]\\
        e^{j2\pi m\Delta f(\frac{m'T}{M}-\tau_p)}e^{j2\pi\nu_p(n'T+\frac{m'T}{M})} & \!\!\mathrm{if}~ m'<\frac{M\tau_p}{T}\\
    \end{array}
\right.
\end{aligned}
\end{equation}
\end{small}%
applying the DFT and using the orthogonal property, the output is given by
\begin{small}
\begin{equation}
\begin{aligned}
&y[n',m']=\frac{1}{\sqrt{M}}\sum_{i=0}^{M-1}y'_{n',i}e^{-j2\pi\frac{m'i}{M}}\\
&=\sum_{p=1}^P\sum_{m=0}^{M-1}\!\!\Big(\Psi^{p,1}_{n',m'}[m]x[n'\!-\!1,m]+\Psi^{p,2}_{n',m'}[m]x[n',m]\Big)
\end{aligned}
\label{modelofdm}
\end{equation}
\end{small}
where
\begin{small}
\begin{equation}
\begin{aligned}
\!\!\!\Psi^{p,1}_{n',m'}[m]&=\frac{1}{\sqrt{M}}h_pe^{j2\pi (\nu_pn'T-m\Delta f\tau_p)}\!\!\sum_{i=0}^{\lfloor\frac{M\tau_p}{T}\rfloor}\!\!e^{j2\pi\frac{(m-m'+\nu_pT)i}{M}}\\
\!\!\!\Psi^{p,2}_{n',m'}[m]&=\frac{1}{\sqrt{M}}h_pe^{j2\pi (\nu_pn'T-m\Delta f\tau_p)}\!\!\!\!\sum_{i=\lceil\frac{M\tau_p}{T}\rceil}^{M-1}\!\!\!e^{j2\pi\frac{(m-m'+\nu_pT)i}{M}}\\
\end{aligned}
\end{equation}
\end{small}%
notations $\lceil\cdot\rceil$ and $\lfloor\cdot\rfloor$ indicates the ceil and floor operation, respectively. Equation \eqref{modelofdm} builds the element-wise relationship between the transmitter and the receiver.
\subsection{Point-to-Point CP-OFDM Transmission}
A more standard modulation is CP-OFDM where a CP is added before each data symbol to avoid ISI and to admit symbol-by-symbol detection, we show that the input-output relation in this case can be simplified. The symbol duration for CP-OFDM is $T'=T_{cp}+T$, where $T_{cp}$ is the CP duration, and is configured to be larger than the maximum multipath delay.
The transmitted OFDM signal in this case should be altered as
\begin{small}
\begin{equation}
\begin{aligned}
f(t)=\sum_{m=0}^{M-1}\sum_{n=0}^{N-1}x[n,m]e^{j2\pi(f_c+m\Delta f)(t-nT')}g_{tx}(t-nT'-T_{cp})
\end{aligned}
\end{equation}
\end{small}%
still sampling $y'(t)$ every $T/M$ seconds in each OFDM symbol, but discard the CP samples to remove the ISI, we obtain
\begin{small}
\begin{equation}
\begin{aligned}
&y'_{n',m'}=y'(t)|_{t=n'T'+T_{cp}+m'T/M}\\
&=\!\sum_{p=1}^{P}\!h_p\!\!\!\sum_{m=0}^{M-1}\!\!x[n',m]e^{j2\pi\nu_p(n'T'+T_{cp}+\frac{m'T}{M})}e^{j2\pi m\Delta f(\frac{m'T}{M}-\tau_p)}
\end{aligned}
\end{equation}
\end{small}%
after discrete Fourier transform, the output is given by
\begin{small}
\begin{equation}
\begin{aligned}
y[n',m']=&\frac{1}{\sqrt{M}}\sum_{i=0}^{M-1}y'_{n',i}e^{-j2\pi\frac{m'i}{M}}\\
=&\frac{1}{\sqrt{M}}\sum_{p=1}^{P}\sum_{m=0}^{M-1}x[n',m]e^{-j2\pi m\Delta f\tau_p}\\
&e^{j2\pi\nu_p(n'T'+T_{cp})}\mathrm{Dir}(m-m'+\nu_pT,M)\\
=&\sum_{p=1}^{P}\sum_{m=0}^{M-1}\Psi^p_{n',m'}[m]x[n',m]
\end{aligned}
\label{ofdm_sig_model}
\end{equation}
\end{small}
where
\begin{small}
\begin{equation}
\begin{aligned}
&\Psi^p_{n',m'}[m]=\\
&\frac{1}{\sqrt{M}}h_pe^{-j2\pi m\Delta f\tau_p}e^{j2\pi\nu_p(n'T'+T_{cp})}\mathrm{Dir}(m-m'+\nu_pT,M)
\end{aligned}
\end{equation}
\end{small}%
here we simplify the expression by using the Dirichlet kernel function $\mathrm{Dir}(\phi,Z)=\sum_{z=0}^{Z-1}e^{j2\pi\phi\frac{z}{Z}}$, the value of which is equal to $Z$ (or $-Z$ depending if $Z$ is even or odd) when $\phi$ is a multiple of $Z$, and is equal to zero for all other integer values of $\phi$.
\subsection{Point-to-Point Zak-OTFS Transmission}
For point-to-point OTFS transmission using Zak receiver, the transmitter first maps symbols $x[k,l]$ to samples $X[n,m]$, from the DD domain to the time-frequency domain using inverse symplectic finite Fourier transform (ISFFT)\cite{RavPhaHon:J18}
\begin{equation}
\begin{aligned}
X[n,m]=\frac{1}{\sqrt{NM}}\sum_{k=0}^{N-1}\sum_{l=0}^{M-1}x[k,l]e^{j2\pi(\frac{nk}{N}-\frac{ml}{M})}
\end{aligned}
\end{equation}
for $n=0,...,N-1$, $m=0,...,M-1$. Next, the time-frequency modulator converts the samples $X[n,m]$ to a continuous time waveform $s(t)$ using shaping pulse $g_{tx}(t)$, i.e.,
\begin{small}
\begin{equation}
\begin{aligned}
f(t)&=\sum_{m=0}^{M-1}\sum_{n=0}^{N-1}X[n,m]e^{j2\pi m\Delta f(t-nT)}g_{tx}(t-nT)\\
&=\frac{1}{\sqrt{NM}}\sum_{k=0}^{N-1}\sum_{l=0}^{M-1}x[k,l]\beta_{k,l}(t)\\
\end{aligned}
\label{otfs_tx_sig}
\end{equation}
\end{small}
\begin{equation}
\begin{aligned}
\beta_{k,l}(t)=\sum_{m=0}^{M-1}\sum_{n=0}^{N-1}g_{tx}(t-nT)e^{j2\pi\frac{nk}{N}}e^{j2\pi m\Delta f(t-\frac{lT}{M})}
\end{aligned}
\end{equation}
where $\beta_{k,l}(t)$ is the time-domain information carrying signal for the $(k,l)$-th information symbol $x[k,l]$. Similar to OFDM, with the aid of DD spread function $h(\tau,\nu)$, the received signal after go through the wireless channel is given by
\begin{equation}
\begin{aligned}
y(t)=\sum\nolimits_{p=1}^{P}h_pf(t-\tau_p)e^{j2\pi\nu_pt}.
\end{aligned}
\label{ytotfs}
\end{equation}

In the receiver, in order to decode the transmitted data, the signal is transformed to the DD domain via Zak transform. For the complex time-continuous signal $y(t)$, the Zak representation of which can be written as\cite{Jan:J88}
\begin{equation}
\begin{aligned}
Z_y(\tau,\nu)=\sqrt{T}\sum\nolimits_{n=-\infty}^{\infty}y(\tau+nT)e^{-j2\pi n\nu T}
\end{aligned}
\label{Zy}
\end{equation}
which relates the DD representation of the signal and its time domain waveform. The received DD domain samples can be obtained directly by sampling the Zak representation of the received time domain signal without going through a time frequency domain transit, i.e.,
\begin{equation}
\begin{aligned}
Y[k',l']=Z_y(\tau=l'T/M,\nu=k'\Delta f/N)
\end{aligned}
\end{equation}
for $k'=0,1,...,N-1,l'=0,1,...,M-1$. The transmitted DD domain information symbols $x[k,l],k=0,1,...,N-1,l=0,1,...,M-1$ are decoded from these received DD domain samples.
\subsection{Point-to-Point Two-Steps OTFS Transmission}
For point-to-point OTFS transmission based on two-steps conversion method, the received time signal $y(t)$ first passes through a matched filter to computes the cross ambiguity function in the following way
\begin{equation}
\begin{aligned}
Y(t,f)=\int g_{rx}^{*}(t'-t)y(t')e^{-j2\pi ft'}dt'
\end{aligned}
\end{equation}
where $g_{rx}(t)$ denotes the received pulse shaping function. By substituting \eqref{ytotfs} and reordering terms, we obtain
\begin{small}
\begin{equation}
\begin{aligned}
&Y(t,f)=\sum_{n'=0}^{N-1}\sum_{m'=0}^{M-1}X[n',m']\int \int h(\tau,\nu)\int g^{*}_{rx}(t'-t)\\
&g_{tx}(t'-\tau-n'T)e^{j2\pi m'\Delta f(t'-\tau)}e^{j2\pi \nu t'}e^{-j2\pi ft'}dt'd\tau d\nu
\end{aligned}
\end{equation}
\end{small}%
the output of the matched filter is obtained by sampling $Y(t,f)$ as
\begin{equation}
\begin{aligned}
Y[n,m]=Y(t,f)|_{t=nT,f=m\Delta f}
\end{aligned}
\end{equation}
therefore, we obtain
\begin{small}
\begin{equation}
\begin{aligned}
Y[n,m]=&\sum_{n',m'}\sum_{p=0}^{P-1}h_pA_{c}((n-n')T-\tau_p,(m-m')\Delta f-\nu_p)\\
&e^{j2\pi\nu_p n'T}e^{j2\pi\nu_p\tau_p}e^{-j2\pi m\Delta f\tau_p}X[n',m'].\\
\end{aligned}
\end{equation}
\end{small}%

By applying the SFFT, we obtain the DD domain representation of the received signal as
\begin{small}
\begin{equation}
\begin{aligned}
Y[k',l']=&\frac{1}{NM}\sum_{n=0}^{N-1}\sum_{m=0}^{M-1}Y[n,m]e^{-j2\pi(\frac{nk'}{N}-\frac{ml'}{M})}\\
=&\sum_{k,l}\sum_{n,m}\sum_{n',m'}\sum_{p=0}^{P-1}\frac{1}{NM}h_pA_{c}\Big((n-n')T-\tau_p,\\
&(m-m')\Delta f-\nu_p\Big)e^{j2\pi\nu_p n'T}e^{j2\pi\nu_p\tau_p}e^{-j2\pi m\Delta f\tau_p}\\
&e^{j2\pi(\frac{n'k}{N}-\frac{m'l}{M})}e^{-j2\pi(\frac{nk'}{N}-\frac{ml'}{M})}x[k,l]
\end{aligned}
\end{equation}
\end{small}%
where $A_c(\tau,\nu)$ is the cross ambiguity function of $g_{tx}(t)$ and $g_{rx}(t)$. Since the received signal $r(t)$ is sampled at time intervals $t'=T/M$, we get
\begin{small}
\begin{equation}
\begin{aligned}
A_{c}\Big((n&-n')T-\tau_p,(m-m')\Delta f-\nu_p\Big)\\
=&\frac{T}{M}\sum_{i=-\infty}^{\infty}g^{*}_{rx}\Big(i\frac{T}{M}-(n-n')T+\tau_p\Big)g_{tx}\Big(i\frac{T}{M}\Big)\\
&e^{-j2\pi[(m-m')\Delta f-\nu_p]\frac{i}{M\Delta f}}
\end{aligned}
\end{equation}
\end{small}%
therefore, we obtain
\begin{small}
\begin{equation}
\begin{aligned}
\!\!\!Y[k',l']=&\sum_{k,l}\frac{x[k,l]}{NM}T\sum_{p=0}^{P-1}h_pe^{j2\pi\nu_p\tau_p}\Big\{\sum_{n,n'}g^{*}_{rx}\Big(l\frac{T}{M}-(n\\
&-n')T+\tau_p\Big)g_{tx}\Big(l\frac{T}{M}\Big)\mathrm{Dir}(l'-l-\tau_pM\Delta f,M)\\
&e^{-j2\pi k'\frac{n}{N}}e^{j2\pi\nu_p\frac{l}{M\Delta f}}e^{j2\pi(k+\nu_pNT)\frac{n'}{N}}\Big\}.
\end{aligned}
\end{equation}
\end{small}%

Consider the default rectangular waveform of $g_{tx}(t)$, and reasonably assume that the maximum possible delay satisfies $\tau<T$, means only the signal of the first preceding symbol duration is involved in the ISI calculation, then the sum $\sum_{n'}$ takes into account only two terms, i.e., $n'=n$ and $n'=n-1$. The indice $l$ involved in the two terms is different since it considers the pulses overlap within the interval $\mathcal{I}_1=[0,M-1-\lceil\tau_pM/T\rceil]$ and $\mathcal{I}_2=[M-1-\lfloor\tau_pM/T\rfloor,M-1]$. After some algebra we can finally obtain
\begin{small}
\begin{equation}
\begin{aligned}
y[k',l']=\sum_{k=0}^{N-1}\sum_{l=0}^{M-1}\sum_{p=1}^{P}\Psi^p_{k',k}[l',l]x[k,l]
\end{aligned}
\end{equation}
\end{small}
where
\begin{small}
\begin{equation}
\begin{aligned}
&\Psi^p_{k',k}[l',l]=h_p\frac{1}{MN}e^{j2\pi\nu_p\tau_p}\mathrm{Dir}(\nu_pNT-k'+k,N)\mathrm{Dir}(l'-\\
&l-\tau_pM\Delta f,M)e^{j2\pi\nu_p\frac{l}{M\Delta f}}\left\{
    \begin{array}{lc}
        1 & \mathrm{if}\quad l\in \mathcal{I}_1, \\
        e^{-j2\pi(\nu_pT+\frac{k}{N})} & \mathrm{if}\quad l\in \mathcal{I}_2.\\
    \end{array}
\right.
\end{aligned}
\label{Htwostepsotfs}
\end{equation}
\end{small}%

\section{CRLB Derivation}
CRLB expresses a lower bound on the variance of unbiased estimators of deterministic parameters. It can be used as a metric to describe the performance limit of OTFS and OFDM systems for sensing. Noteworthy that in order to ensure compatibility with existing OFDM systems, most prior work on OTFS receivers consider a two-step conversion, where the received time-domain signal is firstly converted to the time-frequency domain using an OFDM demodulator, followed by post-processing of this time-frequency signal into DD domain\cite{RavPhaHon:J18}. Recently proposed OTFS modem scheme based on the Zak transform allows the received signal to be converted directly from the time domain to the DD domain thus simplifying the input-output relationship. For either approach, we will then show by CRLB derivation and numerical analysis that they both outperform OFDM in terms of sensing performance.

For CRLB derivation of the two systems, we rephrase the input-output relationship in matrix form as
\begin{equation}
\begin{aligned}
\V{y}=\sum\nolimits_{p=1}^{P}\M{\Psi}^p\V{x}
\end{aligned}
\label{mat_form_sig_model}
\end{equation}
with $\V{y}=[\V{y}^{\mathrm{T}}_0,...,\V{y}^{\mathrm{T}}_{N-1}]^{\mathrm{T}}$ and $\V{x}=[\V{x}^{\mathrm{T}}_0,...,\V{x}^{\mathrm{T}}_{N-1}]^{\mathrm{T}}$ vectors of dimension $NM\times 1$ obtained by stacking the received samples and information symbols, respectively, where $\V{y}_{n}=[y[n,0],y[n,1],...,y[n,M-1]]^{\mathrm{T}}$, $\V{x}_{n}=[x[n,0],x[n,1],...,x[n,M-1]]^{\mathrm{T}}$. The detailed expression of the channel coefficient matrix $\M{\Psi}^p \in \mathbb{C}^{MN\times MN}$ differs for these four signal models, as will be discussed subsequently.
\subsection{CRLB of CP-Free OFDM}
For CP-Free OFDM modulation, based on the element-wise input-output relationship in \eqref{modelofdm}, the channel matrix can be written as
\begin{small}
\begin{equation}
\M{\Psi}^p = \left[\begin{array}{cccc}
\M{\Psi}^{p,2}_{0} &                  &                  &        \\
\M{\Psi}^{p,1}_{1} & \V{\Psi}^{p,2}_{1} &                  &      \\
                 &     \ddots           & \ddots               &  \\
                 &                     & \M{\Psi}^{p,1}_{N-1} & \M{\Psi}^{p,2}_{N-1}\\
\end{array}\right]
\end{equation}
\end{small}%
where
\begin{small}
\begin{equation}
\M{\Psi}^{p,i}_{n'} = \left[\begin{array}{ccc}
\Psi^{p,i}_{n',0}[0] & \cdots   &  \Psi^{p,i}_{n',0}[M-1] \\
\vdots               &  \ddots  &    \vdots              \\
\Psi^{p,i}_{n',M-1}[0] & \cdots   &  \Psi^{p,i}_{n',M-1}[M-1]\\
\end{array}\right]
\end{equation}
\end{small}%
for $i=1,2$.

The partial derivatives w.r.t the magnitude and phase of $h_p$ are straightforward and omitted for the sake of brevity. The derivatives of $\Psi^{p,i}_{n',m'}[m],i={1,2}$ w.r.t $\tau_p$ and $\nu_p$ are more cumbersome and, after some algebra, can be obtained as
\begin{small}
\begin{equation}
\begin{aligned}
\frac{\partial\Psi^{p,1}_{n',m'}[m]}{\partial \tau}=&\frac{1}{\sqrt{M}}h_p(-j2\pi m\Delta f)e^{j2\pi(\nu_pn'T-m\Delta f\tau_p)}\\
&\sum\nolimits_{i=0}^{\lfloor\frac{M\tau_p}{T}\rfloor}e^{j2\pi\frac{(m-m'+\nu_pT)i}{M}}
\end{aligned}
\end{equation}
\end{small}
\begin{small}
\begin{equation}
\begin{aligned}
\frac{\partial\Psi^{p,1}_{n',m'}[m]}{\partial \nu}=&\frac{1}{\sqrt{M}}h_pe^{j2\pi(\nu_pn'T-m\Delta f\tau_p)}\sum\nolimits_{i=0}^{\lfloor\frac{ M\tau_p}{T}\rfloor}\\
&\Big[j2\pi(Ti/M+n'T)\Big]e^{j2\pi\frac{(m-m'+\nu_pT)i}{M}}
\end{aligned}
\end{equation}
\end{small}
\begin{small}
\begin{equation}
\begin{aligned}
\frac{\partial\Psi^{p,2}_{n',m'}[m]}{\partial \tau}=&\frac{1}{\sqrt{M}}h_p(-j2\pi m\Delta f)e^{j2\pi(\nu_pn'T-m\Delta f\tau_p)}\\
&\sum\nolimits_{i=\lceil\frac{M\tau_p}{T}\rceil}^{M-1}e^{j2\pi\frac{(m-m'+\nu_pT)i}{M}}
\end{aligned}
\end{equation}
\end{small}
\begin{small}
\begin{equation}
\begin{aligned}
\frac{\partial\Psi^{p,2}_{n',m'}[m]}{\partial \nu}=&\frac{1}{\sqrt{M}}h_pe^{j2\pi(\nu_pn'T-m\Delta f\tau_p)}\sum\nolimits_{i=\lceil\frac{M\tau_p}{T}\rceil}^{M-1}\\
&\Big[j2\pi(Ti/M+n'T)\Big]e^{j2\pi\frac{(m-m'+\nu_pT)i}{M}}.
\end{aligned}
\end{equation}
\end{small}%
\subsection{CRLB of CP-OFDM}
For CP-OFDM, based on the input-output relationship in \eqref{ofdm_sig_model}, the channel matrix $\M{\Psi}^p$ can be expressed as $\M{\Psi}^p=\mathrm{bdiag}\{\M{\Psi}^p_0,...,\M{\Psi}^p_{N-1}\}$, where
\begin{small}
\begin{equation}
\M{\Psi}_{n'}^p = \left[\begin{array}{ccc}
\Psi^p_{n',0}[0] & \cdots  & \Psi^p_{n',0}[M-1] \\
\vdots & \ddots & \vdots\\
\Psi^p_{n',M-1}[0] & \cdots & \Psi^p_{n',M-1}[M-1]\\
\end{array}\right]
\end{equation}
\end{small}%
and $\mathrm{bdiag}\{\cdot\}$ denotes the diagonalization operation.
The derivatives w.r.t $\tau_p$ and $\nu_p$ can be deduced as
\begin{small}
\begin{equation}
\frac{\partial\Psi^p_{n',m'}[m]}{\partial\tau_p}=[-j2\pi m\Delta f]\Psi^p_{n',m'}[m]
\end{equation}
\end{small}
\begin{small}
\begin{equation}
\begin{aligned}
\frac{\partial\Psi^p_{n',m'}[m]}{\partial\nu_p}=&\frac{1}{\sqrt{M}}h_pe^{-j2\pi m\Delta f\tau_p}e^{j2\pi\nu_p(n'T'+T_{cp})}\sum_{i=0}^{M-1}\\
&\Big[j2\pi(\frac{iT}{M}+n'T'+T_{cp})\Big]e^{j2\pi\frac{i}{M}(m-m'+\nu_pT)}.\\
\end{aligned}
\end{equation}
\end{small}

\subsection{CRLB of Zak-OTFS}
Based on the aforementioned processing flow of the Zak receiver, by substituting the expression of $y(t)$ into \eqref{Zy}, the Zak representation of the received signal is
\begin{small}
\begin{equation}
\begin{aligned}
Z_y(\tau,\nu)=&\sqrt{T}\sum_{n'=-\infty}^{\infty}\sum_{p=0}^{P-1}\sum_{k=0}^{N-1}\sum_{l=0}^{M-1}h_pf(\tau-\tau_p+n'T)\\
&e^{j2\pi\nu_p(\tau+n'T)}e^{-j2\pi n'\nu T}\\
=&\frac{\sqrt{T}}{\sqrt{MN}}\sum_{n'=-\infty}^{\infty}\Big[\sum_{p=0}^{P-1}h_p\sum_{k=0}^{N-1}\sum_{l=0}^{M-1}x[k,l]e^{j2\pi\nu_p(\tau+n'T)}\\
&e^{j2\pi n'\nu T}\beta_{k,l}(\tau-\tau_p+n'T)\Big]\\
=&\frac{\sqrt{T}}{\sqrt{MN}}\sum_{n'=-\infty}^{\infty}\Big[\sum_{p=0}^{P-1}h_p\sum_{k=0}^{N-1}\sum_{l=0}^{M-1}x[k,l]\sum_{m=0}^{M-1}\sum_{n=0}^{N-1}\\
&g(\tau-\tau_p+(n'-n)T)e^{j2\pi k\frac{n}{N}}e^{j2\pi m\Delta f(\tau-\tau_p+n'T-\frac{lT}{M})}\\
&e^{j2\pi\nu_p(\tau+n'T)}e^{j2\pi n'\nu T}\Big]\\
\end{aligned}
\end{equation}
\end{small}
therefore, the sampled received symbol can be given by
\begin{small}
\begin{equation}
\begin{aligned}
\!\!Y[k',l']=Z_y\Big(\frac{l'T}{M},\frac{k'\Delta f}{N}\Big)=\sum_{p=1}^{P}\sum_{k=0}^{N-1}\sum_{l=0}^{M-1}\Psi^p_{k',k}[l',l]x[k,l]
\end{aligned}
\label{otfsmodel}
\end{equation}
\end{small}
\begin{small}
\begin{equation}
\begin{aligned}
\Psi^p_{k',k}[l',l]=&\frac{\sqrt{T}}{\sqrt{MN}}h_p\sum_{m=0}^{M-1}\sum_{n=0}^{N-1}\sum_{n'=-\infty}^{\infty}\!\!\!e^{j2\pi\nu_p(\frac{l'T}{M}+n'T)}\\
&e^{j2\pi m\Delta f(\frac{l'T}{M}-\tau_p+n'T-\frac{lT}{M})}e^{j2\pi \frac{nk}{N}}e^{-j2\pi \frac{k'n'}{N}}\\
&g\big(\frac{l'T}{M}-\tau_p+(n'-n)T\big).\\
\end{aligned}
\end{equation}
\end{small}%

Reasonably assuming that the maximum possible delay satisfies $\tau<T$, that means only the signal of the first preceding symbol is involved in the ISI calculation. Then the sum $\sum_{n'}$ takes into account only two terms, i.e., $n'=n$ in case that $l'$ residues within the interval $\mathcal{I}'_1=[\lceil\tau_pM/T\rceil,M-1]$ and $n'=n+1$ in case that $l$ belongs to interval $\mathcal{I}'_2=[0,\lfloor\tau_pM/T\rfloor]$. We further obtain the channel coefficient as \eqref{hotfs} (shown on the top of the next page), where the expression is simplified by using the Dirichlet kernel function $\mathrm{Dir}(\phi,Z)=\sum_{z=0}^{Z-1}e^{j2\pi\phi\frac{z}{Z}}$, the value of which is equal to $Z$ (or $-Z$ depending if $Z$ is even or odd) when $\phi$ is a multiple of $Z$, and is equal to zero for all other integer values of $\phi$.
\begin{figure*}[ht]
\begin{small}
\begin{equation}
\begin{aligned}
\Psi^p_{k',k}[l',l]=
\left\{
    \begin{array}{lc}
        \frac{1}{\sqrt{MN}}h_pe^{j2\pi\nu_p\frac{l'T}{M}}\mathrm{Dir}(l'-l-\tau_p\Delta f M,M)\mathrm{Dir}(k-k'+\nu_pTN,N) & \mathrm{if}\quad l'\in \mathcal{I}'_1, \\
        \frac{1}{\sqrt{MN}}h_pe^{j2\pi\nu_p\frac{l'T}{M}}e^{j2\pi\nu_pT}e^{-j2\pi\frac{k'}{N}}\mathrm{Dir}(l'-l-\tau_p\Delta f M,M)\mathrm{Dir}(k-k'+\nu_pTN,N) & \mathrm{if}\quad l'\in \mathcal{I}'_2\\
    \end{array}
\right.
\end{aligned}
\label{hotfs}
\vspace{-5mm}
\end{equation}
\end{small}
\end{figure*}

Likewise, the signal model of OTFS can also be expressed in matrix form. Although formally the same as the previous OFDM model, it is worth noting that the input-output relationship of the OFDM system is in the time-frequency domain, while for OTFS it is in the DD domain, and the channel matrix $\M{\Psi}^p$ should be altered as
\begin{small}
\begin{equation}
\M{\Psi}^p = \left[\begin{array}{ccc}
\Psi^p_{0,0} & \cdots  & \Psi^p_{0,N-1} \\
\vdots & \ddots & \vdots\\
\Psi^p_{N-1,0} & \cdots & \Psi^p_{N-1,N-1}\\
\end{array}\right]
\end{equation}
\end{small}
\begin{small}
\begin{equation}
\M{\Psi}^p_{k',k} = \left[\begin{array}{ccc}
\Psi^p_{k',k}[0,0] & \cdots  & \Psi^p_{k',k}[0,M-1] \\
\vdots & \ddots & \vdots\\
\Psi^p_{k',k}[M-1,0] & \cdots & \Psi^p_{k',k}[M-1,M-1].\\
\end{array}\right]
\end{equation}
\end{small}%

For Zak-OTFS, the partial derivatives of $\Psi^p_{k',k}[l',l]$ w.r.t $\tau_p$ and $\nu_p$ can be obtained as \eqref{detau} and \eqref{denu}, respectively.
\subsection{CRLB of Two-Steps OTFS}
Whereas for OTFS modulation based on the two-step conversion, the derivatives of $\Psi^p_{k',k}[l',l]$ in \eqref{Htwostepsotfs} w.r.t $\tau_p$ and $\nu_p$ can be obtained as
\begin{small}
\begin{equation}
\begin{aligned}
&\frac{\partial\Psi^{p}_{k',k}[l',l]}{\partial \tau_p}=\frac{j2\pi}{NM}h_p\mathrm{Dir}(\nu_pNT-k'+k,N)e^{j2\pi\nu_p(\tau_p+\frac{l}{M\Delta f})}\!\!\\
&\sum_{m=0}^{M-1}(\nu_p-m\Delta f)e^{j2\pi(l'-l-\tau_pM\Delta f)\frac{m}{M}}\left\{\!\!\!
    \begin{array}{lc}
        1 & \!\!\!\!\mathrm{if}~ l\in \mathcal{I}_1, \\
        e^{-j2\pi(\nu_pT+\frac{k}{N})} & \!\!\!\!\mathrm{if}~ l\in \mathcal{I}_2\\
    \end{array}
\right.
\end{aligned}
\end{equation}
\end{small}
\begin{small}
\begin{equation}
\begin{aligned}
&\frac{\partial\Psi^p_{k',k}[l',l]}{\partial \nu_p}=\frac{j2\pi}{NM}h_p\mathrm{Dir}(l'-l-\tau_pM\Delta f,M)\\
&e^{j2\pi\nu_p(\tau_p+\frac{l}{M\Delta f})}\sum_{n}e^{j2\pi(\nu_pNT-k'+k)\frac{n}{N}}\\
&\left\{
    \begin{array}{lc}
        \frac{l}{M\Delta f}+nT+\tau_p & \mathrm{if}~ l\in \mathcal{I}_1, \\
        (\frac{l}{M\Delta f}+nT+\tau_p-T)e^{-j2\pi(\nu_pT+\frac{k}{N})} & \mathrm{if}~ l\in \mathcal{I}_2.\\
    \end{array}
\right.
\end{aligned}
\end{equation}
\end{small}

The CRLB of time delays and Doppler shifts follows by filling the Fisher information matrix (FIM) with the derivatives deduced above. FIM is defined as\cite{Sen:J93}
\begin{equation}
\begin{aligned}
\M{J}(\V{\theta})=\mathbb{E}\Big\{\big[\frac{\partial}{\partial\V{\theta}}\mathrm{log}\;p(\V{\V{y},\theta})\big]\big[\frac{\partial}{\partial\V{\theta}}\mathrm{log}\;p(\V{\V{y},\theta})\big]^{\mathrm{T}}\Big\}
\end{aligned}
\label{FIMdefinition}
\end{equation}
where the vector $\V{\theta}\in \mathbb{R}^{4P}$ consists of all parameters of the $P$ paths. Specifically, for path $p$, it includes $\tau_p$ and $\nu_p$, as well as the amplitude and phase of $h_p$. $p(\V{y},\V{\theta})$ is the likelihood function of $\V{y}$ when $\V{\theta}$ is observed.  Assume that the received signal is interfered by an additive white Gaussian noise $\V{n}\sim\mathcal{CN}(\V{0},\M{\Sigma})$, where $\M{\Sigma}$ is the covariance matrix, therefore we have $\V{y}\sim\mathcal{CN}(\M{\Psi}\V{x},\M{\Sigma})$. It can be obtained that the logarithm probability function of $\V{y}$ w.r.t $\V{\theta}$ is
\begin{equation}
\begin{aligned}
\Lambda&=\mathrm{log}\;p(\V{y},\V{\theta})=-(\V{y}-\M{\Psi}\V{x})^{\mathrm{H}}\M{\Sigma}^{-1}(\V{y}-\M{\Psi}\V{x})-C
\end{aligned}
\label{logf}
\end{equation}
where $C$ is a constant term that is not concerned. Based on \eqref{logf} we have
\begin{small}
\begin{equation}
\begin{aligned}
\frac{\partial \Lambda}{\partial \theta_i}=-(\V{y}-\M{\Psi}\V{x})^{\mathrm{H}}\M{\Sigma}^{-1}\frac{\partial \M{\Psi}}{\partial \theta_i}\V{x}-(\frac{\partial \M{\Psi}}{\partial \theta_i}\V{x})^{\mathrm{H}}\M{\Sigma}^{-1}(\V{y}-\M{\Psi}\V{x})
\end{aligned}
\end{equation}
\end{small}%
since $\mathbb{E}\{\V{y}-\M{\Psi}\V{x}\}=0$, it can be got that $\mathbb{E}\{\frac{\partial \Lambda}{\partial \theta_i}\}=0, \thinspace \forall i$. Therefore, we can obtain
\begin{equation}
\begin{aligned}
\mathbb{E}\Big\{\frac{\partial^2 \Lambda}{\partial \theta_i \partial\theta^*_j}\Big\}=-2\mathfrak{Re}\Big\{\big(\frac{\partial \M{\Psi}}{\partial \theta_j}\V{x}\big)^{\mathrm{H}}\M{\Sigma}^{-1}\frac{\partial \M{\Psi}}{\partial \theta_i}\V{x}\Big\}
\end{aligned}
\end{equation}
hence the FIM $\M{J}(\V{\theta})$ can be obtained as
\begin{equation}
\left\{
\begin{aligned}
\M{J}(\V{\theta})&=[J_{i,j}]_{1\le i,j\le 4P}\\
J_{i,j}&=-\mathbb{E}(\frac{\partial^2 f}{\partial \theta_i \partial\theta^*_j})
\end{aligned}
\right.
\end{equation}
the MSE of any unbiased estimator $\hat{\V{\theta}}$ satisfies
\begin{equation}
\begin{aligned}
\mathbb{E}\{(\hat{\V{\theta}}-\V{\theta})(\hat{\V{\theta}}-\V{\theta})^{\mathrm{T}}\}\succeq \M{J}^{-1}(\V{\theta})
\end{aligned}
\end{equation}
where the main diagonal elements of $\M{J}^{-1}(\V{\theta})$ are the desired CRLBs for the unknown parameters.

\begin{figure*}[ht]
\begin{small}
\begin{equation}
\begin{aligned}
\!\!\frac{\partial\Psi^p_{k',k}[l',l]}{\partial \tau_p}=
\left\{
    \begin{array}{lc}
       \!\!\!\!\! \frac{1}{\sqrt{MN}}h_pe^{j2\pi\nu_p\frac{l'T}{M}}\mathrm{Dir}(k-k'+\nu_pTN,N)\sum_{m=0}^{M-1}[-j2\pi m\Delta f]e^{j2\pi m\frac{l'-l}{M}}e^{j2\pi m\Delta f\tau_p} & \!\!\!\!\mathrm{if}~ l'\in \mathcal{I}'_1, \\
        \!\!\!\!\!\frac{1}{\sqrt{MN}}h_pe^{j2\pi\nu_p\frac{l'T}{M}}e^{j2\pi \nu_p T}e^{-j2\pi\frac{k'}{N}}\mathrm{Dir}(k-k'+\nu_pTN,N)\sum_{m=0}^{M-1}[-j2\pi m\Delta f]e^{j2\pi m\frac{l'-l}{M}}e^{j2\pi m\Delta f\tau_p} & \!\!\!\!\mathrm{if}~l'\in \mathcal{I}'_2\\
    \end{array}
\right.
\end{aligned}
\label{detau}
\vspace{-5mm}
\end{equation}
\end{small}%
\end{figure*}
\begin{figure*}[ht]
\begin{small}
\begin{equation}
\begin{aligned}
\!\!\!\frac{\partial\Psi^p_{k',k}[l',l]}{\partial \nu_p}=
\left\{
    \begin{array}{lc}
        \!\!\!\!\!
        \frac{1}{\sqrt{MN}}h_p\mathrm{Dir}(l'-l-\tau_p\Delta f M,M)\sum_{n=0}^{N-1}[j2\pi(\frac{l'T}{M}+nT)]e^{j2\pi n\frac{(k-k')}{N}}e^{j2\pi(\frac{l'T}{M}+nT)\nu_p} & \!\!\!\!\!\mathrm{if}~ l'\in \mathcal{I}'_1, \\
        \!\!\!\!\!\frac{1}{\sqrt{MN}}h_pe^{-j2\pi\frac{k'}{N}}\mathrm{Dir}(l'-l-\tau_p\Delta f M,M)\sum_{n=0}^{N-1}[j2\pi(\frac{l'T}{M}+nT+T)]e^{j2\pi n\frac{(k-k')}{N}}e^{j2\pi(\frac{l'T}{M}+nT+T)\nu_p} & \!\!\!\!\!\mathrm{if}~ l'\in \mathcal{I}'_2\\
    \end{array}
\right.
\end{aligned}
\label{denu}
\vspace{-6mm}
\end{equation}
\end{small}
\end{figure*}
\section{numerical results}
\begin{figure}[t]
\centering
\subfigure[CRLB of delay versus SNR]{\includegraphics[width=1.72in,height=1.4in]{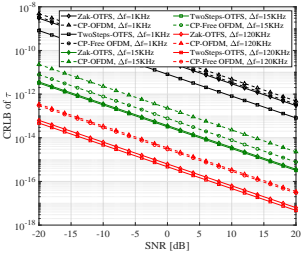}}
\subfigure[CRLB of Doppler versus SNR]{\includegraphics[width=1.72in,height=1.4in]{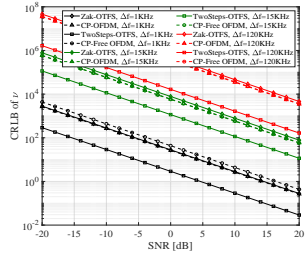}}
\caption{CRLB illustration of $\tau$ and $\nu$ versus SNR. Key configurations includes $N=M=12$, $f_c=3$GHz, $\tau=3.33\times10^{-6}$s and $\nu=500$Hz. }
\label{CRLB_SNR_SCS}
\end{figure}

\begin{figure}[t]
\centering
\subfigure[CRLB of delay versus MN]{\includegraphics[width=1.72in,height=1.4in]{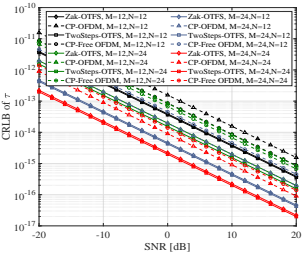}}
\subfigure[CRLB of Doppler versus MN]{\includegraphics[width=1.72in,height=1.4in]{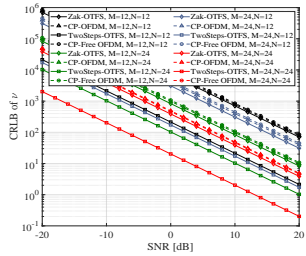}}
\caption{CRLB illustration of $\tau$ and $\nu$ versus the size of $M$ and $N$. Other parameters are the same as Fig. \ref{CRLB_SNR_SCS}. }
\label{CRLB_MN}
\end{figure}

\begin{figure}[t]
\centering
\subfigure[CRLB of delay versus SNR]{\includegraphics[width=1.72in,height=1.4in]{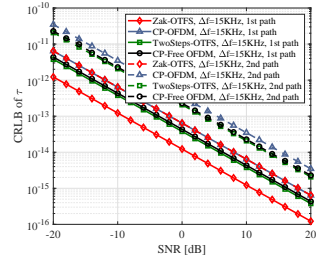}}
\subfigure[CRLB of Doppler versus SNR]{\includegraphics[width=1.72in,height=1.4in]{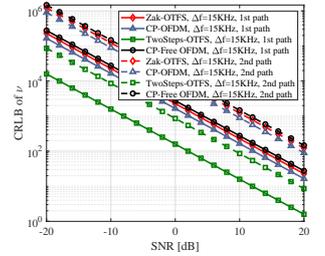}}
\caption{CRLB illustration of $\tau$ and $\nu$ versus SNR in case of two paths channel. Key parameters includes $M=N=12$, $h_1=0.7e^{j\pi/3}$, $h_2=0.3e^{j3\pi/4}$, $\tau_1=3.33\times10^{-6}$s, $\nu_1=500$Hz, $\tau_2=5\times10^{-6}$s and $\nu_2=2.5$KHz. }
\label{CRLB_twopath}
\end{figure}

In this section, we evaluate the CRLB of OFDM and OTFS modulations as functions of signal-to-noise ratio (SNR), SCS, as well as PRB size. Firstly, we consider a single path, SCS$=15$KHz, and fix the PRB size to be $M=N=12$. The CRLB of delay and Doppler shift versus SNR is presented in Fig. \ref{CRLB_SNR_SCS}, which shows that the bound of delay estimation decreases with increasing SCS, and vice versa for that of Doppler shift, because a large SCS implies a wide bandwidth and short time duration. This reveals that it is feasible to satisfy the sensing requirements of various applications by adjusting the SCS. Next, Fig. \ref{CRLB_MN} shows the impact of PRB size on the CRLB. As expected, more PRBs help improve sensing accuracy, which is still essentially a result of changes in bandwidth and time duration in the case of a given SCS. Besides, given PRB size, diverse delay and Doppler estimation performance can be met by tuning M and N combinations. Finally, we present the CRLBs of $\tau$'s and $\nu$'s in the multipath case in Fig. \ref{CRLB_twopath}, where the first path has lower delay and Doppler CRLB than the second path due to higher signal strength. Overall, OTFS has a dominant advantage in delay and Doppler estimation compared to OFDM, where the Zak-OTFS and the Two-Steps OTFS outperform other schemes in terms of delay and Doppler shift estimation, respectively. The bound for CP-Free OFDM or CP-OFDM are comparable.

\section{Conclusion}
\indent In this paper, we investigated the sensing performance of OTFS and OFDM modulation systems based on the constructed point-to-point transmission models. For OFDM, we consider the ideal CP-Free and the practical CP-added models, respectively, while for OTFS, both the Zak transform based and the Two-Step conversion based receivers are considered. The CRLBs for the delay and Doppler shift under these four models are then derived, respectively. Numerical results validate that OTFS outperforms OFDM for sensing, while the Zak-OTFS and the Two-Steps OTFS outperform other schemes in terms of delay and Doppler shift estimation, respectively. In addition, the analysis of the effect of SCS and PRB size on the CRLB provides guidance on satisfying the sensing requirements of various scenarios through flexible parameter tuning.


\vspace{12pt}

\bibliographystyle{IEEEtran}

\bibliography{ref}

\begin{thebibliography}{10}
\providecommand{\url}[1]{#1}
\csname url@samestyle\endcsname
\providecommand{\newblock}{\relax}
\providecommand{\bibinfo}[2]{#2}
\providecommand{\BIBentrySTDinterwordspacing}{\spaceskip=0pt\relax}
\providecommand{\BIBentryALTinterwordstretchfactor}{4}
\providecommand{\BIBentryALTinterwordspacing}{\spaceskip=\fontdimen2\font plus
\BIBentryALTinterwordstretchfactor\fontdimen3\font minus
  \fontdimen4\font\relax}
\providecommand{\BIBforeignlanguage}[2]{{%
\expandafter\ifx\csname l@#1\endcsname\relax
\typeout{** WARNING: IEEEtran.bst: No hyphenation pattern has been}%
\typeout{** loaded for the language `#1'. Using the pattern for}%
\typeout{** the default language instead.}%
\else
\language=\csname l@#1\endcsname
\fi
#2}}
\providecommand{\BIBdecl}{\relax}
\BIBdecl

\bibitem{LiuHuaLi:J22}
A.~Liu, Z.~Huang, M.~Li, Y.~Wan, W.~Li, T.~X. Han, C.~Liu, R.~Du, D.~K.~P. Tan,
  J.~Lu, Y.~Shen, F.~Colone, and K.~Chetty, ``A survey on fundamental limits of
  integrated sensing and communication,'' \emph{IEEE Communications Surveys \&
  Tutorials}, vol.~24, no.~2, pp. 994--1034, 2022.

\bibitem{StuBarMcl:J04}
G.~Stuber, J.~Barry, S.~McLaughlin, Y.~Li, M.~Ingram, and T.~Pratt, ``Broadband
  {MIMO-OFDM} wireless communications,'' \emph{Proceedings of the IEEE},
  vol.~92, no.~2, pp. 271--294, 2004.

\bibitem{WanSheMaz:J14}
T.~Wang, Y.~Shen, S.~Mazuelas, H.~Shin, and M.~Z. Win, ``On {OFDM} ranging
  accuracy in multipath channels,'' \emph{IEEE Systems Journal}, vol.~8, no.~1,
  pp. 104--114, 2014.

\bibitem{StuWie:J11}
C.~Sturm and W.~Wiesbeck, ``Waveform design and signal processing aspects for
  fusion of wireless communications and {R}adar sensing,'' \emph{Proceedings of
  the IEEE}, vol.~99, no.~7, pp. 1236--1259, 2011.

\bibitem{RavPhaHon:J18}
P.~Raviteja, K.~T. Phan, Y.~Hong, and E.~Viterbo, ``Interference cancellation
  and iterative detection for orthogonal time frequency space modulation,''
  \emph{IEEE Transactions on Wireless Communications}, vol.~17, no.~10, pp.
  6501--6515, 2018.

\bibitem{GauKobCai:J20}
L.~Gaudio, M.~Kobayashi, G.~Caire, and G.~Colavolpe, ``On the effectiveness of
  {OTFS} for joint {R}adar parameter estimation and communication,'' \emph{IEEE
  Transactions on Wireless Communications}, vol.~19, no.~9, pp. 5951--5965,
  2020.

\bibitem{RavPhaHon:C19}
P.~Raviteja, K.~T. Phan, Y.~Hong, and E.~Viterbo, ``Orthogonal time frequency
  space ({OTFS}) modulation based {Radar} system,'' in \emph{2019 IEEE Radar
  Conference (RadarConf)}, 2019, pp. 1--6.

\bibitem{Sen:J93}
S.~K. Sengijpta, ``Fundamentals of statistical signal processing: Estimation
  theory,'' \emph{Technometrics}, vol.~37, no.~4, pp. 465--466, 1993.

\bibitem{Moh:J21}
S.~K. Mohammed, ``Derivation of otfs modulation from first principles,''
  \emph{IEEE Transactions on Vehicular Technology}, vol.~70, no.~8, pp.
  7619--7636, Mar. 2021.

\bibitem{Bel:J63}
P.~Bello, ``Characterization of randomly time-variant linear channels,''
  \emph{IEEE Transactions on Communications Systems}, vol.~11, no.~4, pp.
  360--393, 1963.

\bibitem{Jan:J88}
A.~J. Janssen, ``The {Z}ak transform: a signal transform for sampled
  time-continuous signals.'' \emph{Philips Journal of Research}, vol.~43,
  no.~1, pp. 23--69, 1988.

\end{thebibliography}

\end{document}